\documentclass[aps,epjc,twocolumn,floatfix,superscriptaddress,nofootinbib,showpacs,showkeys]{revtex4}
\usepackage{graphicx}
\usepackage{dcolumn}  % Align table columns on decimal point
\usepackage{bm}       % bold math
\usepackage{palatino}
\usepackage{epsfig}
\begin{document}
\title{Average formation lengths of baryons and antibaryons in string model
}
\author{L.~Grigoryan \\
Yerevan Physics Institute, Br.Alikhanian 2, 375036 Yerevan, Armenia }
%%%%%%%%%%%%%%%  begin abstract  %%%%%%%%%%%%%%%%%%%%%%%%%%%%%%%.ps&
\begin {abstract}
\hspace*{1em}
In this work it is continued the investigation of the space-time scales of the
hadronization process in the framework of string model. The average formation
lengths of several widely using species of baryons (antibaryons) such as $p$ 
($\bar{p}$), $n$ ($\bar{n}$), $\Delta$ ($\bar{\Delta}$), $\Lambda$ ($\bar{\Lambda}$)
and $\Sigma$ ($\bar{\Sigma}$) are studied. It is shown that they depend from
electrical charges or, more precise, from quark contents of the hadrons.
In particular, the average formation lengths of positively charged hadrons, for
example protons, are considerably larger than of their negatively charged 
antiparticles, antiprotons. This statement is fulfilled
for all nuclear targets and any value of the Bjorken scaling variable $x_{Bj}$.
The main mechanism is direct production. Additional production mechanism in result
of decay of resonances gives small contribution. It is shown that the average
formation lengths of protons (antiprotons) are slowly rising (decreasing) functions
of $x_{Bj}$, the ones of neutrons and antineutrons are slowly decreasing functions
of $x_{Bj}$. The shape and behavior of average formation lengths for baryons 
qualitatively coincide with the ones for pseudoscalar mesons obtained earlier.
\end {abstract}
\maketitle
%%%%%%%%%%%%%%%  end abstract  %%%%%%%%%%%%%%%%%%%%%%%%%%%%%%%
%%%%%%%%%%%%%%%  begin \section{Introduction}  %%%%%%%%%%%%%%%%%%%%%%%%%
\section{Introduction}
\normalsize
\hspace*{1em}
In any hard process the initial interaction takes place between partons
which then turn into the final hadrons by means of hadronization process.
The space-time evolution of the hadronization process, despite on its
importance, is known relatively little. In particular in
Refs.~\cite{chmaj,Bi_Gyul} the average formation lengths of high-energy
hadrons was studied based on the Lund model of hadronization~\cite{lund}.
The ambiguity in the concept of formation length for composite particles
was pointed out. Two different formation lengths were defined and their
distributions calculated. The results were compared with the data which
allowed to choice suitable form for average formation length.\\
\hspace*{1em}
In the Ref.~\cite{grignew} the investigation of the space-time scales of
the hadronization process was continued for the concrete case of
pseudoscalar mesons, produced in semi-inclusive deep inelastic scattering
(DIS). It was shown that the average formation lengths of these hadrons
depend from their electrical charges. In particular the average formation
lengths of positively charged mesons are larger than of negatively charged
ones. This statement was verified for $z$ (the fraction of the virtual
photon energy transferred to the detected hadron) in the current
fragmentation region, for cases of different scaling functions, for all
nuclear targets and any value of the Bjorken scaling variable $x_{Bj}$.
In all cases, the main mechanism was the direct production of pseudoscalar
mesons. Including in consideration the additional mechanism of pseudoscalar
mesons production in result of decay of resonances, leaded to the decrease
of average formation lengths. It was shown that the average formation
lengths of positively (negatively) charged mesons were slowly rising
(decreasing) functions of $x_{Bj}$.\\
\hspace*{1em}
The investigation of average formation lengths of baryons and antibaryons
is the next step in the study of space-time structure of hadronization
process. The mechanism for meson production follows rather naturally from
the simple picture of a meson as a short piece of string between $q$ and
$\bar{q}$ endpoints. There is no unique recipe to generalize this picture
to baryons. In the framework of Lund model~\cite{lund,eden,seostrand} the baryon
in DIS can be produced in three scenarios: (i) diquark scenario; (ii) simple
popcorn scenario; (iii) advanced popcorn scenario. (i) Diquark picture.
Baryon production may, in its simplest form, be obtained by assuming that
any flavor $q_{i}$, produced from color field of string, could represent
either a quark or an antidiquark in a color triplet state. Then the same
basic formalism can be used as in case of meson production, supplemented
with the probability to produce various diquark pairs. In this simple picture
the baryon and the antibaryon are produced as neighbours in rank
\footnote{Rank denotes the order in which the hadrons are produced, counting
from the string end.}
in a string breakup.\\
\hspace*{1em}
The experimental data indicate that occasionally one or a few mesons may be
produced in between the baryon and the antibaryon ($B\bar{B}$) along the
string. This fact was used for development the so called popcorn model. The
popcorn model is a more general framework for baryon production, in which
diquarks as such are never produced, but rather baryons appear from the
successive production of several $q_{i}\bar{q_{i}}$ pairs. It is evidently
the density and the size of the color fluctuations which determine the
properties of the $B\bar{B}$ production process. The density determines the
rate of baryon production but in case the fluctuations are large on the scale
of the meson masses it is possible that one or more mesons are produced
between the $B\bar{B}$-pair. Taking into account the uncertainty principle we
can estimate the value of the color fluctuations. It turn out that there is a
fast fall-off with the size of the space-time regions inside which color
fluctuations may occur. Therefore, in a model of this kind, $B\bar{B}$
produced in pair are basically either nearest neighbours or next nearest
neighbours in rank. (ii) Simple popcorn. In this model it is assumed that at
most one meson could be produced between the baryon and antibaryon. It is
assumed that $B\bar{B}$ and $BM\bar{B}$ (with additional meson between $B$
and $\bar{B}$) configurations occur with equal probability. (iii) Advanced
popcorn. It is assumed that several mesons could be produced between the
baryon and antibaryon. This model has more complicated set of parameters.
In this work, for the sake of simplicity, the diquark picture will be used.\\
In section 2 the theoretical framework is briefly presented. In section 
3 the obtained results are presented and discussed. The section 4 contains 
conclusions.
%%%%%%%%%%%%%%%%%%%%%%%%%%%%%%%%%%%
%%%%%%%%%%%%%%%  end \section{Introduction}  %%%%%%%%%%%%%%%%%%%%%%%%%
%%%%%%%%%%%  begin \section{Theoretical framework}  %%%%%%%%%%%
\section{Theoretical framework}
\hspace*{1em}
In Refs.~\cite{akopov2,akopov3,grig} it was shown that a ratio of
multiplicities for the nucleus and deuterium can be presented in the form of
a function of single variable which has the physical meaning of the formation
length (time) of the hadron. This scaling was verified, for the case of
charged pions, by HERMES experiment~\cite{airap3}. Now HERMES experiment
prepares two-dimensional analysis of nuclear attenuation data.\\
\hspace*{1em}
In the string model, for the construction of fragmentation functions, the
scaling function $f(z)$ is introduced (see, for instance,
Refs.~\cite{lund,andersson}). It is defined by the condition that $f(z)dz$ is
the probability that the first hierarchy (rank 1) primary hadron carries away
the fraction of energy $z$ of the initial string. We use symmetric Lund
scaling function~\cite{lund,seostrand} for calculations:
\begin{eqnarray}
f(z) = Nz^{-1}(1 - z)^{a}exp(-bm_{\perp}^{2}/z) ,
\end{eqnarray}
where $a$ and $b$ are parameters of model,
$m_{\perp}=\sqrt{m_{h}^{2}+p_{\perp}^{2}}$ is the transverse mass of final
hadron, $N$ is normalization factor.\\
\hspace*{1em}
In the further study we will use the average value of the formation length
defined as $L_{c}^{h} = <l_c>$.\\
\hspace*{1em}
The consideration is convenient to begin from $L_{c}^{h}$ direct,
$L_{c}^{h(dir)}$, which takes into account the direct production of hadrons:
\begin{eqnarray}
{L_{c}^{h(dir)} = \int_0^{\infty} ldlD_{c}^{h}(L,z,l)/\int_0^{\infty} 
dlD_{c}^{h}(L,z,l)}
\hspace{0.3cm},
\end{eqnarray}
where $L = \nu/\kappa$ is the full hadronization length, $\kappa$ is the
string tension (string constant), $D_{c}^{h}(L, z, l)$ is the distribution of
the constituent formation length $l$ of hadrons carrying fractional energy $z$.
%%% new text %%%
\begin{eqnarray}
\nonumber
D_{c}^{h}(L,z,l) = \Big(C_{p1}^{h}f(z)\delta(l-L+zL)+
\end{eqnarray}
\begin{eqnarray}
C_{p2}^{h}\sum_{i=2}^{n}D_{ci}^{h}(L,z,l)\Big)\theta(l)\theta(L-zL-l)
\hspace{0.3cm}.
\end{eqnarray}
The functions $C_{p1}^{h}$ and $C_{p2}^{h}$ are the probabilities that in
electroproduction process on proton target the valence quark compositions for
leading (rank 1) and subleading (rank 2) hadrons will be obtained. Similar
functions were obtained in~\cite{akopov1} for more general case of nuclear
targets. In eq.(3) $\delta$- and $\theta$-functions arise as a consequence
of energy conservation law. The functions $D_{ci}^{h}(L,z,l)$ are
distributions of the constituent formation length $l$ of the rank $i$ hadrons
carrying fractional energy $z$. For calculation of distribution functions we
used recursion equation from Ref.~\cite{Bi_Gyul}.\\
\hspace*{1em}
The simple form of $f(z)$ for standard Lund model allows to sum the sequence
of produced hadrons over all ranks ($n=\infty$). The analytic expression for
the distribution function in this case was presented in~\cite{grignew}.\\
\hspace*{1em}
Unfortunately, in case of more complicated scaling function presented in
eq.(1) the analytic summation of the sequence of produced hadrons over all
ranks is impossible. Therefore, we limited ourself by $n=10$ in eq.(3).\\
\hspace*{1em}
In the Ref.~\cite{grignew} the essential contributions in the spectra of
pseudoscalar mesons from the decays of vector mesons were obtained. Now,
using the same formalism we will calculate the contribution of baryonic
resonances in the average formation lengths of baryons.\\
\hspace*{1em}
The distribution function of the constituent formation length $l$ of the
daughter hadron $h$ which arises in result of decay of parent resonance $R$
and carries away the fractional energy $z$ is denoted $D_c^{R/h}(L, z, l)$.
It can be computed from the convolution integral:
\begin{eqnarray}
\nonumber
D_c^{R/h}(L, z, l) = d^{R/h}
\int_{z_{down}^{R/h}}^{z_{up}^{R/h}} \frac{dz'}{z'}D_{c}^{R}(L,z',l)\times
\end{eqnarray}
\begin{eqnarray}
f^{R/h}\Big(\frac{z}{z'}\Big) 
\hspace{0.3cm},
\end{eqnarray}
where $z_{up}^{R/h} = min(1,z/z_{min}^{R/h})$ and
$z_{down}^{R/h} = min(1,z/z_{max}^{R/h})$, $z_{max}^{R/h}$ ($z_{min}^{R/h}$)
is maximal (minimal) fraction of the energy of parent resonance, which can be
carried away by the daughter baryon.\\
\hspace*{1em}
Let us consider the two-body isotropic decay of resonance $R$,
$R \to h_{1}h_{2}$, and denote the energy and momentum of the daughter hadron
$h$ ($h = h_{1}$ or $h_{2}$), in the rest system of resonance, $E_{h}^{(0)}$
and $p_{h}^{(0)}$, respectively. In the coordinate system where resonance has
energy and momentum equal $E_R$ and $p_R$,
\begin{eqnarray}
z_{max}^{R/h} = \frac{1}{m_{R}}\Big(E_{h}^{(0)} +
\frac{p_R}{E_R}p_{h}^{(0)}\Big)
\hspace{0.3cm},
\end{eqnarray} 
\begin{eqnarray}
z_{min}^{R/h} = \frac{1}{m_{R}}\Big(E_{h}^{(0)} -
\frac{p_R}{E_R}p_{h}^{(0)}\Big)
\hspace{0.3cm},
\end{eqnarray}
where $m_R$ is the mass of resonance $R$. In the laboratory (fixed target)
system the resonance usually fastly moves, i.e. $p_R/E_R \to 1$.\\
\hspace*{1em}
The constants $d^{R/h}$ can be found from the branching ratios in the decay
process $R \to h$. We will present their values for interesting for us cases
below.\\
\hspace*{1em}
The distributions $f^{R/h}(z)$ are determined from the decay process of the
resonance R, with momentum $p$ into the hadron $h$ with momentum $zp$. We
assume that the momentum $p$ is much larger than the masses and the transverse
momenta involved.\\
\hspace*{1em}
In analogy with eq.(2) we can write the expression for the average value of
the formation length $L_{c}^{R/h}$ for the daughter baryon $h$ produced in
result of decay of the parent resonance $R$ in form:
\begin{eqnarray}
{L_{c}^{R/h} = \int_0^{\infty} ldlD_{c}^{R/h}(L,z,l)/
\int_0^{\infty} dlD_{c}^{R/h}(L,z,l)}
\hspace{0.3cm}.
\end{eqnarray}\\
Here it is need to give some explanations. We can formally consider
$L_{c}^{R/h}$ as the formation length of daughter baryon $h$ for two reasons:
(i) the parent resonance and daughter hadron are the hadrons of the same rank,
which have common constituent quark; (ii) beginning from this distance the
chain consisting from prehadron, resonance and final baryon $h$ interacts (in
nuclear medium) with hadronic cross sections.\\
\hspace*{1em}
The general formula for $L_{c}^{h}$ for the case when a few resonances
contribute can be written in form:
\begin{eqnarray}
\nonumber
L_{c}^{h} = \int_0^{\infty} ldl\Big(\alpha_{B}D_c^{h}(L,z,l) + 
\alpha_{R}\sum_{R}D_c^{R/h}(L,z,l)\Big)/
\end{eqnarray}
\begin{eqnarray}
\int_0^{\infty} dl\Big(\alpha_{B}D_c^{h}(L,z,l) +             
\alpha_{R}\sum_{R}D_c^{R/h}(L,z,l)\Big)
\hspace{0.15cm},
\end{eqnarray}
where $\alpha_{B}$ ($\alpha_{R}$) is the probability that $qqq$ system turns
into baryon (baryonic resonance). For $\Delta$ and $\Sigma$ resonances,
taking into account the decuplet/octet suppression and the extra
$\Sigma/\Lambda$ suppression following from the mass
differences~\cite{eden,chliap}, the condition $\alpha_{R}$=$\alpha_{B}$=0.5
is used.\\
\hspace*{1em}
Let us now discuss the details of model, which are necessary for calculations.
We will consider several species of widely using baryons (antibaryons) such as
$p$ ($\bar{p}$), $n$ ($\bar{n}$), $\Delta$ ($\bar{\Delta}$), $\Lambda$ 
($\bar{\Lambda}$) and $\Sigma$ ($\bar{\Sigma}$) electroproduced on proton,
neutron and nuclear targets. The scaling function $f(z)$ in eq.(1) has two free
parameters~\cite{seostrand} $a=0.8$, $b=0.58 GeV^{-2}$. Next parameter, which is 
necessary for the calculations in the framework of string model is the string
tension. It was fixed at a static value determined by the Regge trajectory 
slope~\cite{seostrand,kappa}
\begin{eqnarray}
\kappa = 1/(2\pi\alpha'_R) = 1 GeV/fm \hspace{0.3cm}.
\end{eqnarray}
\hspace*{1em}
Now let us turn to the functions $C_{p1}^{h}$ and $C_{p2}^{h}$, which have the
physical meaning of the probabilities to produce on proton target hadron h of
first and second ranks, respectively. For pseudoscalar mesons they were
presented in Ref.~\cite{grignew}. For baryons these functions have more
complicate structure, therefore it is convenient to present here final
expressions which were obtained after small calculations. For protons and
antiprotons of first and second ranks they have the form
\begin{eqnarray}
\nonumber
C_{p1}^{p} = \frac{\frac{4}{9}u(x_{Bj},Q^2){\cdot}1.05 + 
\frac{1}{9}d(x_{Bj},Q^2){\cdot}0.1}
{\sum_{q=u,d,s} e_{q}^{2}(q(x_{Bj},Q^2) + \bar{q}(x_{Bj},Q^2))} \gamma_{qq0} ,
\end{eqnarray}
\begin{eqnarray}
\nonumber
C_{p1}^{\bar{p}} = \frac{\frac{4}{9}\bar{u}(x_{Bj},Q^2){\cdot}1.05 + 
\frac{1}{9}\bar{d}(x_{Bj},Q^2){\cdot}0.1}
{\sum_{q=u,d,s} e_{q}^{2}(q(x_{Bj},Q^2) + \bar{q}(x_{Bj},Q^2))} \gamma_{qq0} ,
\end{eqnarray}
\begin{eqnarray}
\nonumber
C_{p2}^{p} = C_{p2}^{\bar{p}} = 1.15\gamma_{q}\gamma_{qq0} .
\end{eqnarray}
The coefficients for first rank neutron (antineutron) $C_{p1}^{n}$ 
($C_{p1}^{\bar{n}}$)
can be obtained from corresponding expressions for proton (antiproton) by means of 
changing distribution
functions in nominators ($u,d$)$\to$($d,u$) (($\bar{u}$,$\bar{d}$)$\to$ 
($\bar{d}$,$\bar{u}$)); coefficients for second rank are equal to them for proton 
$C_{p2}^{n}$ = $C_{p2}^{\bar{n}}$ = $C_{p2}^{p}$. For $\Lambda$ and $\bar{\Lambda}$
they are:
\begin{eqnarray}
\nonumber
C_{p1}^{\Lambda} = \frac{(\frac{4}{9}u + \frac{1}{9}d)\gamma_{qs}{\cdot}0.5 + 
\frac{1}{9}s\gamma_{qq0}}
{\sum_{q=u,d,s} e_{q}^{2}(q(x_{Bj},Q^2) + \bar{q}(x_{Bj},Q^2))}  ,
\end{eqnarray}
\begin{eqnarray}
\nonumber
C_{p1}^{\bar{\Lambda}} = \frac{(\frac{4}{9}\bar{u} + 
\frac{1}{9}\bar{d})\gamma_{qs}{\cdot}0.5 +
\frac{1}{9}\bar{s}\gamma_{qq0}}
{\sum_{q=u,d,s} e_{q}^{2}(q(x_{Bj},Q^2) + \bar{q}(x_{Bj},Q^2))}  ,
\end{eqnarray}
\begin{eqnarray}
\nonumber
C_{p2}^{\Lambda} = C_{p2}^{\bar{\Lambda}} = \gamma_{q} \gamma_{qs} + \gamma_{s} 
\gamma_{qq0} .
\end{eqnarray}
%%%
For $\Delta$ resonances these coefficients are:
\begin{eqnarray}
\nonumber
C_{p1}^{\Delta^{++}} = \frac{\frac{4}{9}u(x_{Bj},Q^2)}
{\sum_{q=u,d,s} e_{q}^{2}(q(x_{Bj},Q^2) + \bar{q}(x_{Bj},Q^2))} \gamma_{qq1} ,
\end{eqnarray}
%%%
\begin{eqnarray}
\nonumber
C_{p1}^{\Delta^{+}} = \frac{\frac{4}{9}u(x_{Bj},Q^2){\cdot}\frac{2}{3} +
\frac{1}{9}d(x_{Bj},Q^2){\cdot}\frac{1}{3}}
{\sum_{q=u,d,s} e_{q}^{2}(q(x_{Bj},Q^2) + \bar{q}(x_{Bj},Q^2))} \gamma_{qq1} ,
\end{eqnarray}
%%%
\begin{eqnarray}
\nonumber
C_{p1}^{\Delta^{0}} = \frac{\frac{4}{9}u(x_{Bj},Q^2){\cdot}\frac{1}{3} +
\frac{1}{9}d(x_{Bj},Q^2){\cdot}\frac{2}{3}}
{\sum_{q=u,d,s} e_{q}^{2}(q(x_{Bj},Q^2) + \bar{q}(x_{Bj},Q^2))} \gamma_{qq1} ,
\end{eqnarray}
%%%
\begin{eqnarray}
\nonumber
C_{p1}^{\Delta^{-}} = \frac{\frac{1}{9}d(x_{Bj},Q^2)}
{\sum_{q=u,d,s} e_{q}^{2}(q(x_{Bj},Q^2) + \bar{q}(x_{Bj},Q^2))} \gamma_{qq1} ,
\end{eqnarray}
%%%
\begin{eqnarray}
\nonumber
C_{p2}^{\Delta^{++}} = C_{p2}^{\Delta^{+}} = C_{p2}^{\Delta^{0}} =
C_{p2}^{\Delta^{-}} = \gamma_{q}\gamma_{qq1} ,
\end{eqnarray}
%%%
where $x_{Bj} = \frac{Q^{2}}{2m_{p}\nu}$ is the Bjorken's
scaling variable;
$Q^{2} = -q^{2}$, where $q$ is the 4-momentum of virtual photon; $m_{p}$ is
the proton mass; $q(x_{Bj},Q^2) (\bar{q}(x_{Bj},Q^2))$, where $q=u,d,s$ are
quark (antiquark) distribution functions for proton. Easily to see, that
functions $C_{pn}^{h}$ for hadrons of higher rank ($n > 2$) coincide with
ones for second rank hadron $C_{pn}^{h} \equiv C_{p2}^{h}$. This fact was
already used for construction of eq.(3). For neutron and nuclear targets more
general functions $C_{fi}^{h} (i = 1,2)$ from~\cite{akopov1} are used. 
Similar functions can be obtained for $\bar{\Delta}$, $\Sigma$ and
$\bar{\Sigma}$ resonances.\\
%%%%%%%%%%%  end \section{Theoretical framework}  %%%%%%%%%%%%%%%
%%%%%%%%%%%%%%%  begin \section{Results and Discussion}  %%%%%%%%%%%
\section{Results and Discussion}
\normalsize
\hspace*{1em}
All calculations were performed at fixed value of $\nu$ equal $10 GeV$.
Calculations of $z$ - dependence were performed at fixed value of $Q^{2}$
equal $2.5 GeV^{2}$, which correspond to $x_{Bj} \approx 0.133$. The
parameterizations for quark (antiquark) distributions in proton in
approximation of leading order were taken from~\cite{grv}. We assume,
that new $q\bar{q}$ pairs are $u\bar{u}$ with probability $\gamma_{u}$,
$d\bar{d}$ with probability $\gamma_{d}$ and $s\bar{s}$ with probability
$\gamma_{s}$. It is followed from isospin symmetry that
$\gamma_{u} = \gamma_{d} = \gamma_{q}$.
In the diquark scenario we need also in probabilities for production in
color field of string diquark-antidiquark pairs with different contents:
(i) diquark-antidiquark pairs of light quarks with spin (S) and isospin (I)
S=I=0 $\gamma_{ud0}$=$\gamma_{qq0}$; (ii) diquark-antidiquark pairs of light
quarks with S=I=1 $\gamma_{uu1}$=$\gamma_{ud1}$=$\gamma_{dd1}$=$\gamma_{qq1}$;
(iii) diquark-antidiquark pairs containing strange quark (antiquark)
$\gamma_{us}$=$\gamma_{ds}$=$\gamma_{qs}$. We use the connections between
different quantities and set of values for $\gamma$ from Lund
model~\cite{seostrand} $\gamma_{qq1}$=0.15$\gamma_{qq0}$; $\gamma_{qs}$=
0.12$\gamma_{qq0}$; $\gamma_{qq0}$=0.1$\gamma_{q}$.
$\gamma_{u} : \gamma_{d} : \gamma_{s} = 1 : 1 : 0.3$.\\
\hspace*{1em}
We take into account that part of baryons can be produced from decay
of baryonic resonances. As a possible sources of $p$, $n$ ($\bar{p}$,
$\bar{n}$) and $\Lambda$ ($\bar{\Lambda}$) baryons we consider $\Delta$ 
($\bar{\Delta}$) and $\Sigma$ ($\bar{\Sigma}$) baryonic resonances,
respectively. The contributions of other resonances are neglected.\\
%- begin FIGURE 1 ---------------------------------------------------
\begin{figure}[!htb]
\begin{center}
\epsfxsize=8.cm
\epsfbox{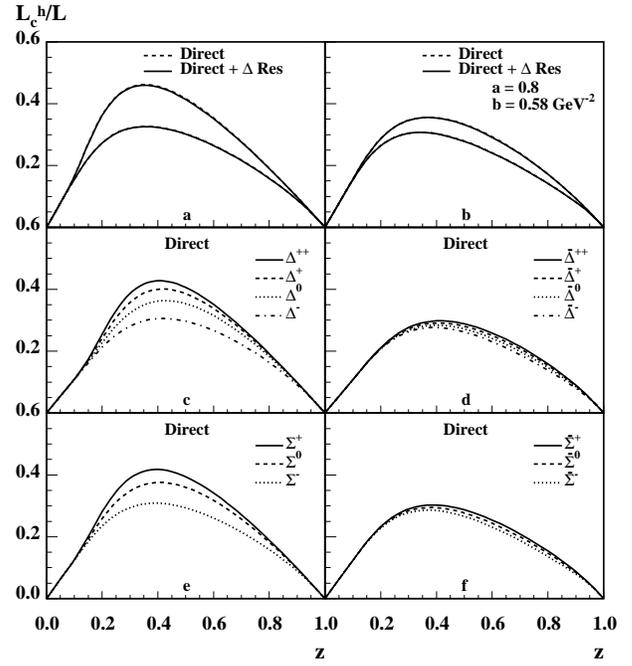}
\end{center}
\caption{\label{xx1}
{\it Average formation lengths for electroproduction of baryons and
antibaryons on proton target, normalized on $L$, are presented as a
functions of $z$. On panel a the protons and antiprotons are presented.
The contributions of direct protons (dashed curves) as well as of the
sum of direct and produced from decay of $\Delta$ resonances protons
(solid curves) are presented. Upper curves represent the average
formation lengths for protons and lower curves the same for antiprotons. 
On panel b the results for neutron and antineutron in the same approach
are presented. On other panels the results for baryonic resonances in
case of direct production are presented. On panel c the results for
$\Delta$ resonances, on panel d for $\bar{\Delta}$ resonances, on panel e
for $\Sigma$ resonances, and on panel f for $\bar{\Sigma}$ resonances are
presented. The results for symmetric Lund model are presented. The
parameters of model also are presented.
}}
\end{figure}
%- end FIGURE 1 ----------------------------------------------------- 
%- begin FIGURE 2 ---------------------------------------------------
\begin{figure}[!htb]
\begin{center}
\epsfxsize=7.5cm
\epsfbox{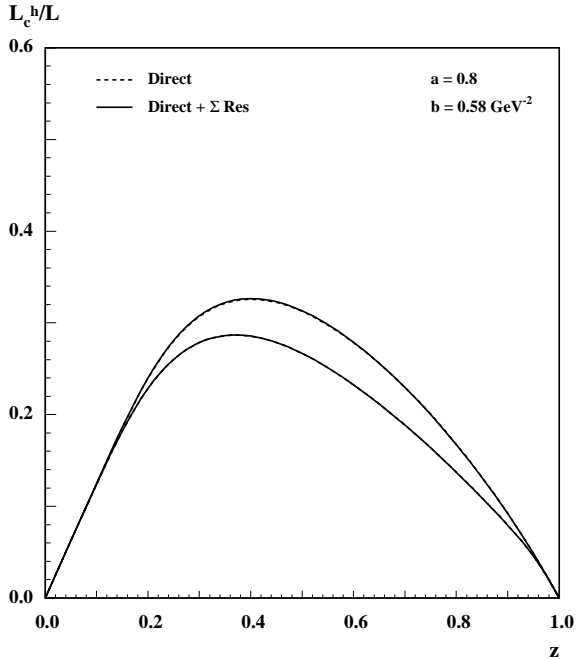}
\end{center}
\caption{\label{xx2}
{\it Average formation lengths for electroproduction of $\Lambda$ and
$\bar{\Lambda}$ on proton target, normalized on $L$, are presented as a
functions of $z$. The contributions of direct hadrons as well as of the
sum of direct and produced from decay of $\Sigma^0$ ($\bar{\Sigma}^0$)
resonances hadrons are presented. Upper curves represent formation
lengths of $\Lambda$ and lower curves of $\bar{\Lambda}$.
}}
\end{figure}
%- end FIGURE 2 -----------------------------------------------------
\hspace*{1em}
The decay distributions $f^{R/h}$ are determined from the decay process of
the resonance R, with momentum p into the hadron h with momentum zp.\\
\hspace*{1em}
In Refs.~\cite{grignew,andersson} they were presented for
case of pseudoscalar mesons, here we will use similar functions for baryons.
The common expression $f^{R/h}(z) = 1/(z_{max}^{R/h}-z_{min}^{R/h})$ 
for all using baryonic resonances will be used. 
The values of $z_{max}^{R/h}$ and $z_{min}^{R/h}$ it is easily to obtain from
eqs.(5) and (6).\\
\hspace*{1em}
For protons we have
$d^{\Delta^{++}/p} = \frac{1}{2}$, $d^{\Delta^{+}/p} = \frac{1}{3}$,
$d^{\Delta^{0}/p} = \frac{1}{6}$; for neutrons $d^{\Delta^{+}/n} = \frac{1}{6}$,
$d^{\Delta^{0}/n} = \frac{1}{3}$, $d^{\Delta^{-}/n} = \frac{1}{2}$; for $\Lambda$
from $\Sigma^{0}$ decay we have $d^{\Sigma^{0}/\Lambda} = \frac{1}{2}$.\\
%%% FIG 1 & 2 %%%
\hspace*{1em}
In Fig.1 the
average formation lengths for electroproduction of baryons and antibaryons
on proton target, normalized on $L$, are presented as a functions of $z$.
On panel a the average formation lengths for
protons and antiprotons are presented. The contributions of direct
protons (dashed curves) as well as of the sum of direct and produced from decay
of $\Delta$ resonances protons (solid curves) are presented. Upper curves
represent the average formation lengths for protons and lower curves the same
for antiprotons. On panel b the results for neutron and antineutron in the same 
approach are presented. On other panels (c, d, e, f) the results for baryonic 
resonances in the approach of direct production are presented. On panel c the
results for $\Delta$ resonances, on panel d for $\bar{\Delta}$ resonances, on
panel e for $\Sigma$ resonances, and on panel f for $\bar{\Sigma}$ resonances
are presented. All results were obtained in the framework of symmetric Lund model.
In Fig.2 the average formation lengths
for electroproduction of $\Lambda$ and $\bar{\Lambda}$ on proton target,
normalized on $L$, are presented as a functions of $z$. The contributions of
directly produced hadrons as well as of the sum of direct and produced from
decay of $\Sigma^0$ ($\bar{\Sigma}^0$)resonances hadrons are presented. Upper
curves represent formation lengths of $\Lambda$ and lower curves of
$\bar{\Lambda}$. The first observation which can be made from Figs.1 and 2 is
that the contribution of the resonances is small, so small that practically
does not change result. The second observation is that for baryons having
several charge states ($\Delta$ and $\Sigma$) there is following rule: larger
charge corresponds larger average formation length. The third observation is
that all antibaryons independent from charge and other quantum numbers have
practically the same average formation lengths.\\
%%% end of text Fig.1 & 2 %%%
\hspace*{1em}
We already discussed in~\cite{grignew} why the average formation lengths of
positively charged hadrons are larger than of negatively charged ones. It
happens due to the large probability to knock out $u$ quark in result of DIS
(even in case of neutron target). The knocked out quark enter in the composition
of leading hadron, which has maximal formation length. From Figs.1 and 2 easely
to see that all antibaryons, independent from electrical charge and other
quantum numbers, have approximately the same average formation lengths. The
cause of this is that all antibaryons consist from "sea" partons, antiquarks,
and can be, with large probability, hadrons of second or higher ranks. We can
compare antibaryons with $K^{-}$ meson, which has average formation length
smaller than other pseudoscalar mesons, because it is constructed from "sea"
quarks only and practically can not be leading hadron, whereas other pseudoscalar 
mesons can be leading hadrons due to $u$ or $d$ quarks entering in their
composition.
%- begin FIGURE 3 ---------------------------------------------------
\begin{figure}[!htb]
\begin{center}
\epsfxsize=8.cm
\epsfbox{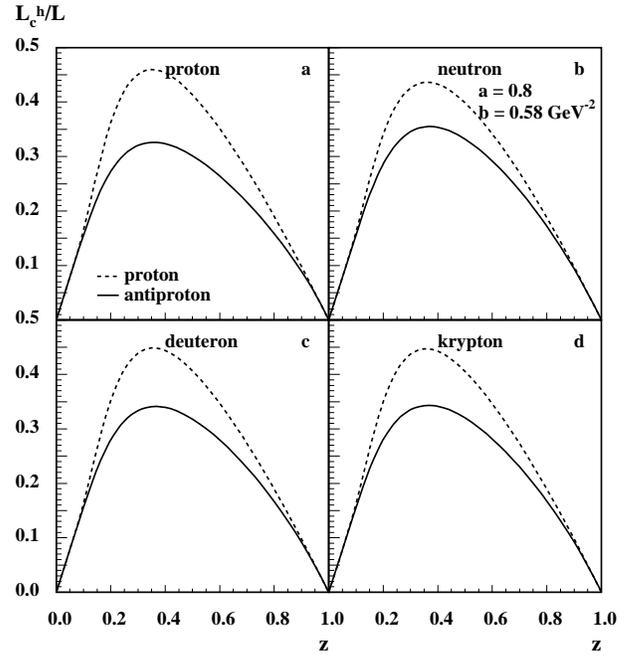}
\end{center} 
\caption{\label{xx3}
{\it Average formation lengths for electroproduction of protons and antiprotons
on different targets, normalized on $L$,
as a functions of $z$.
}}
\end{figure}
%- end FIGURE 3 -----------------------------------------------------
We would like to point out here one interesting feature of baryonic
resonances. Let us calculate from eqs.(5) and (6) the quantities $z_{min}^{R/h}$
and $z_{max}^{R/h}$ for decays $\Delta \to N\pi$ and $\Sigma \to \Lambda \gamma$.
We obtain for first case $z_{min}^{R/h}$=0.6, $z_{max}^{R/h}$=0.97 and for second
case $z_{min}^{R/h}$=0.88, $z_{max}^{R/h}$=1, which means that in case of baryons
integration over $z^{'}$ in eq.(4) is performed in narrow enough region, i.e.
contribution of resonances does not distort the distribution of formation lengths
of hadrons. For comparison, we would like to remind~\cite{grignew}, that for decay
$\rho \to \pi \pi$ $z_{min}^{R/h}$=0.035, $z_{max}^{R/h}$=0.965, which leads to
the distortion of pions spectra.
%- begin FIGURE 4 ---------------------------------------------------
\begin{figure}[!htb]
\begin{center}
\epsfxsize=8.cm
\epsfbox{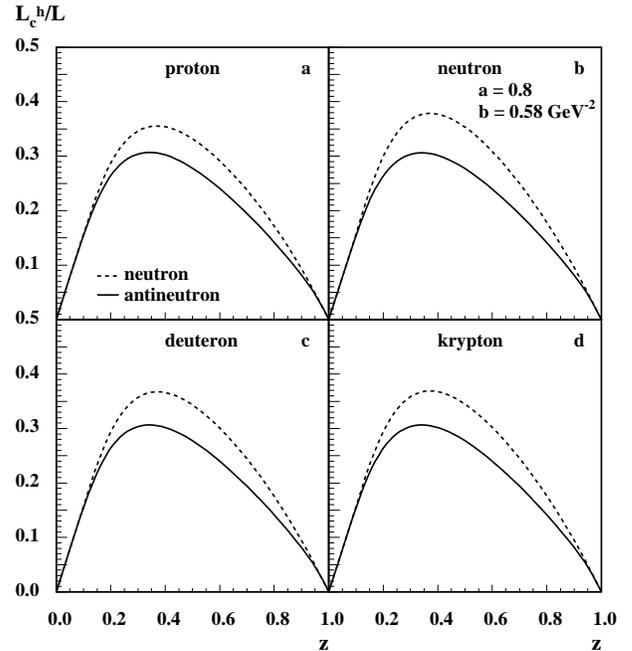}
%\epsfbox{fig2.eps}
\end{center}
\caption{\label{xx4}
{\it Average formation lengths for electroproduction of neutrons and antineutrons
on different targets, normalized on $L$,
as a functions of $z$.
}}
\end{figure}
%- end FIGURE 4 -----------------------------------------------------
%- begin FIGURE 5 ---------------------------------------------------
\begin{figure}[!htb]
\begin{center}  
\epsfxsize=8.cm
\epsfbox{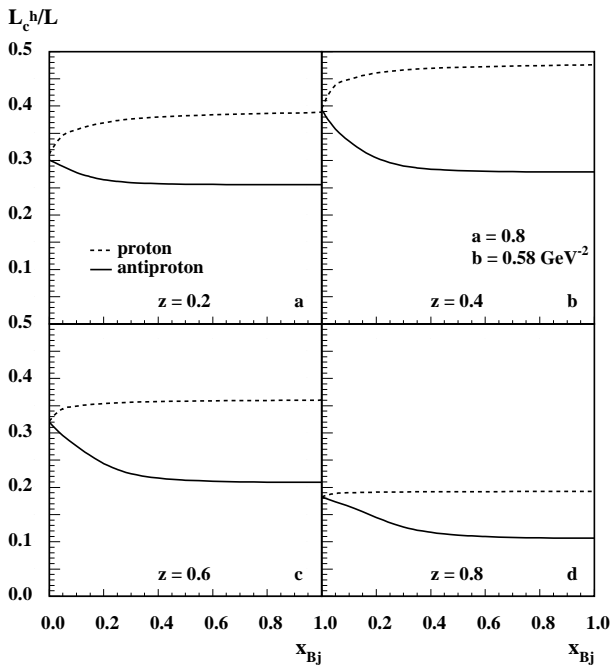}
\end{center}  
\caption{\label{xx5}
{\it Average formation lengths for electroproduction of protons and antiprotons
on proton target, normalized on $L$, as a functions
of $x_{Bj}$.
}}
\end{figure}
%- end FIGURE 5 -----------------------------------------------------
%- begin FIGURE 6 ---------------------------------------------------
\begin{figure}[!htb]
\begin{center}
\epsfxsize=8.cm
\epsfbox{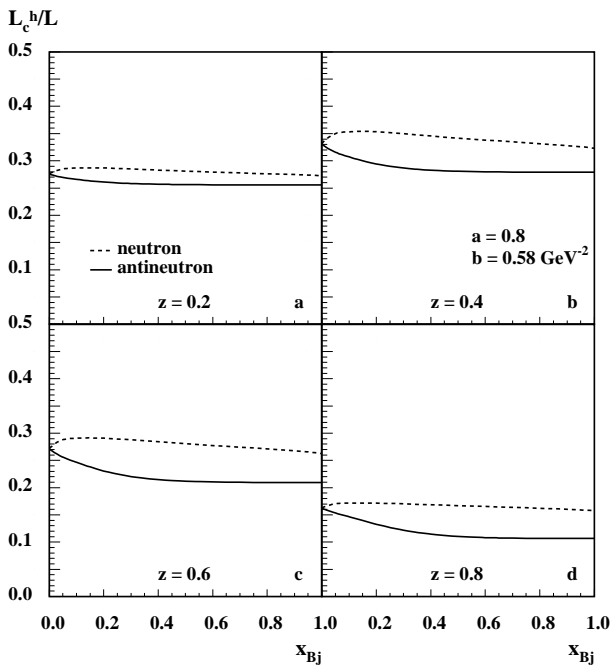}
\end{center}
\caption{\label{xx6}
{\it Average formation lengths for electroproduction of neutrons and antineutrons
on proton target, normalized on $L$, as a functions
of $x_{Bj}$. 
}}
\end{figure}
%- end FIGURE 6 -----------------------------------------------------
It is worth to note, that results for deuteron coincide, in our approach,
with results for any nuclei with $Z = N$, where $Z$ ($N$) is number of protons
(neutrons).
Average formation lengths of hadrons on krypton nucleus, which has essential
excess of neutrons, do not differ considerably from the ones on nuclei with
$Z = N$.
%%% FIG 3 & 4 %%%
In Fig.3 the
average formation lengths for electroproduction of protons and antiprotons
on different targets, normalized on $L$, as a functions of $z$ are presented.
In Fig.4 the average formation lengths for electroproduction of neutrons and 
antineutrons on different targets, normalized on $L$, as a functions of $z$
are presented. From Figs.3 and 4 we have interesting information, that average 
formation lengths of protons (neutrons) reaches maximal value on proton
(neutron) target, which is easily to explain because when the kinds of target
hadron and final hadron coincide, final hadron has maximal chance be leading.
Another information, which is not so obvious as in the previous case, is that 
antiproton has minimal average formation length on proton target (may be it is 
connected with large denominator in function $C_{p1}^{\bar{p}}$).
%%% end of FIG 3 & 4 %%%
%%% FIG 5 & 6 %%%
In Fig.5 the average formation lengths for electroproduction of protons and
antiprotons on proton target, normalized on $L$, as a functions of $x_{Bj}$
are presented. In Fig.6 the same as in Fig.5 for the case of neutrons and
antineutrons are presented. The average formation lengths of protons
(antiprotons) are slowly rising (decreasing) functions of $x_{Bj}$. They
differ significantly at middle $z$, i.e. at middle $z$ antiprotons will
attenuate in nuclei significantly stronger than protons. The average formation
lengths of neutrons (antineutrons) both are slowly decreasing functions of
$x_{Bj}$, their values are close enough for all values of $z$ and for all
region of $x_{Bj}$.
%%% end of FIG 5 & 6 %%%
We obtained, for the first time
\footnote{We would like to note, that in~\cite{accardi} the attempt was made to
obtain average formation lengths for proton and antiproton in the rough version
of standard Lund model. Comparison is shown, that they essentially differ from
our result (Fig.1a).}
, the average formation lengths for different baryons (antibaryons) in the 
eletroproduction process on proton, neutron, deuteron and krypton targets, in
the framework of symmetric Lund model. It is worth to note, that results for
deuteron coincide, in our approach, with results for any nuclei with $Z = N$,
where $Z$ ($N$) is number of protons (neutrons). Average formation lengths of
baryons on krypton nucleus, which has essential excess of neutrons, do not
differ considerably from the ones on nuclei with $Z = N$.  
%%%%%%%%
%%%%%%%%%%%%%%%  end \section{Results and Discussion}  %%%%%%%%%%%

%%%%%%%%%%%%%%%  begin \section{Conclusions}  %%%%%%%%%%%
\section{Conclusions}
%%% New text %%%
Main conclusions are: (i) the average formation lengths of several widely using
species of baryons (antibaryons) such as $p$ ($\bar{p}$), $n$ ($\bar{n}$), $\Delta$ 
($\bar{\Delta}$), $\Lambda$ ($\bar{\Lambda}$) and $\Sigma$ ($\bar{\Sigma}$) for
the case of symmetric Lund model are obtained for the first time; (ii) the average 
formation lengths of baryons and antibaryons produced in semi-inclusive deep
inelastic scattering of leptons on different targets, depend from their electrical
charges or, more precise, from their quark contents; (iii) the contribution of
$\Delta$ ($\bar{\Delta}$) resonances in case of protons (antiprotons) and neutrons 
(antineutrons), and $\Sigma$ ($\bar{\Sigma}$)resonances in case of $\Lambda$ 
($\bar{\Lambda}$) are considered. It is obtained that their contributions are
small, i.e. in case of baryons, production from resonances is essentially weaker
than in case of mesons; (iv) the average formation lengths of protons (antiprotons)
are slowly rising (decreasing) functions of $x_{Bj}$, the average formation
lengths of neutrons and antineutrons are slowly decreasing functions of $x_{Bj}$;
(v) the shape and behavior of average formation lengths for baryons qualitatively 
coincide with the ones for pseudoscalar mesons obtained earlier~\cite{grignew}.\\
%%% End of New text %%%
\hspace*{1em}
It is worth to note that in string model the formation length of the leading
(rank 1) hadron $l_{c1} = (1 - z)\nu/\kappa$ does not depend from type of process, 
kinds of hadron and target. Therefore, the dependence of obtained results from the
type of process, kinds of targets and observed hadrons is mainly due to presence
of higher rank hadrons.\\
\hspace*{1em}
Which sizes can reach the average formation length? At fixed $x_{Bj}$ it is 
proportional to $\nu$. Consequently it will rise with $\nu$ and can reach sizes
much larger than nuclear sizes at very high energies.\\
\hspace*{1em}
At present the hadronization in nuclear medium is widely studied both
experimentally and theoretically. It is well known, that there is nuclear
attenuation of final hadrons. Unfortunately it does not clear, which is the true 
mechanism of such attenuation: final state interactions of prehadrons and
hadrons in nucleus (absorption mechanism); or gluon bremmstrahlung of partons
(produced in DIS) in nuclear medium, whereas hadronization takes place far
beyond nucleus (energy loss mechanism). We hope, that results obtained in the 
previous~\cite{grignew} and this works can be useful for the understanding of this
problem.
%%%%%%%%%%%%%%%  end \section{Conclusions}  %%%%%%%%%%%
%%%%%%%%%%%%%%%%%%%%%%%%%%%%%%%%%%%%%%%%%%%%%%%%%%%%%%%%%%%%%%%%%%%%
%--- begin BIBLIOGRAPHY -------------------

%--- end BIBLIOGRAPHY ---------------------
\end{document}